\documentclass[sigconf]{acmart}
\usepackage{graphicx} 
\usepackage{caption}
\usepackage{subcaption}
\usepackage{tabularx}
\usepackage{booktabs}
\newcolumntype{Y}{>{\raggedright\arraybackslash}X}
\usepackage{textcomp}
\usepackage{xspace} 
\usepackage{url}
\newcommand{\name}{\textit{OriFeel}\xspace}
\usepackage{makecell}

\title{\name: Origami-Inspired Tactile Feedback via Surface Folding}

\author{Shubham Rohal}
\email{srohal@ucmerced.edu}
\affiliation{%
  \institution{University of California, Merced}
  \streetaddress{}
  \city{Merced}
  \state{California}
  \country{USA}
}

\author{Dong Yoon Lee}
\email{dlee267@ucmerced.edu}
\affiliation{%
  \institution{University of California, Merced}
  \streetaddress{}
  \city{Merced}
  \state{California}
  \country{USA}
}

\author{Shijia Pan}
\email{span24@ucmerced.edu}
\affiliation{%
  \institution{University of California, Merced}
  \streetaddress{}
  \city{Merced}
  \state{California}
  \country{USA}
}

\copyrightyear{2026}
\acmYear{2026}
\setcopyright{cc}
\setcctype{by}
\acmConference[UbiComp Companion '26]{Companion of the 2026 ACM International Joint Conference on Pervasive and Ubiquitous Computing}{October 11--15, 2026}{Shanghai, China}
\acmBooktitle{Companion of the 2026 ACM International Joint Conference on Pervasive and Ubiquitous Computing (UbiComp Companion '26), October 11--15, 2026, Shanghai, China}
\acmDOI{10.1145/3798063.3839150}
\acmISBN{979-8-4007-2533-3/2026/10}


\begin{document}

\begin{abstract}
People passively interact with ambient surfaces such as tables, chair backs, and armrests throughout daily life, making them natural candidates for ubiquitous tactile interfaces. However, transforming these everyday surfaces into practical haptic interfaces remains challenging. Existing solutions typically rely on dense arrays of actuators, resulting in bulky hardware and high power consumption that limit their integration into ambient objects.
We present \name, a structure-driven tactile interface that leverages a Miura-ori folding mechanism to distribute actuation across multiple interconnected folding units.
Rather than mapping actuators to individual contact points, \name uses embedded cables to exploit structural interconnections, distributing actuation across the surface and enabling spatially controllable tactile output across multiple folding units.
We implement prototypes using rigid and soft materials and characterize their ability to distribute actuation across interconnected units.
Our results demonstrate the feasibility of structure-driven tactile feedback via coordinated folding of compliant origami surface structures, providing a practical foundation for compact, scalable ambient haptic interfaces.



\end{abstract}





\maketitle

\section{Introduction}
Can surface folding provide a practical path toward ubiquitous haptic feedback on ambient objects?
As interactive surfaces continue to move beyond traditional screens, everyday objects are increasingly envisioned as ambient interfaces. 
Since people naturally maintain passive contact with objects such as tables, chairs, and armrests throughout the day, these ambient surfaces present an underexplored opportunity for delivering unobtrusive tactile feedback.
However, enabling these surfaces to provide compelling haptic feedback remains challenging due to the following reasons. 
First, generating expressive feedback requires densely deployed actuators. 
Second, the bulky, rigid structure of conventional actuators makes them difficult to integrate into thin, deformable, or foldable surfaces.
While dense arrays of vibration motors, motorized pins, and bulky pneumatic pumps are highly effective for handheld devices and virtual reality applications, they do not translate well to passive ambient surfaces due to several physical constraints, including their \textbf{large physical footprint}, limited compatibility with \textbf{elastic surfaces}, and the need for a \textbf{dedicated} actuator beneath each feedback location.

This paper rethinks how to generate tactile feedback to enable ubiquitous haptic interfaces by exploring an alternative structure-driven approach.
By transforming the surface itself into a medium for physical feedback, we explore techniques that enable controlled, predictable folding.
Building on this insight, we present \name, a surface augmentation mechanism that uses an origami structure to generate tactile feedback as normal pressure.
Unlike existing surface haptic interfaces, which rely on a one-to-one actuator-to-feedback spatial mapping, \name decouples actuation from feedback generation. This architecture allows actuators to be placed independently of the feedback location, resulting in (1) \textbf{a compact actuator footprint} that integrates seamlessly into curved and deformable surfaces, (2) \textbf{programmable surface deformation} through compliant origami folding, and (3) \textbf{one-to-many actuation}, where the interconnected origami structure distributes motion to multiple tactile locations from a single actuator.
Therefore, the contribution of this work includes
\begin{itemize}
\item We introduce \name, a surface augmentation mechanism that generates localized normal-pressure feedback through programmable origami folding.
\item We propose a decoupled actuation architecture that separates actuation from feedback generation, enabling compact actuator placement and integration into elastic surfaces.
\item We demonstrate the feasibility of \name with a prototype implementation adopting compliant Miura-Ori structures. 
\end{itemize}

\section{Related Work}\label{sec:related}
We introduce relevant haptic feedback solutions, origami-based design in sensing and robotics, and explain the gap this work fills.


\subsection{Haptic Interfaces}
This section summarizes the state-of-the-art haptic feedback solutions in four categories: vibrotactile, shape-changing, pressure, and kinesthetic interfaces.

\paragraph{Vibration-based Tactile Feedback}\label{sec:related-1}
Vibrotactile interfaces provide tactile feedback by modulating vibration signals in frequency and amplitude~\cite{park2020augmenting, Emgin_2019}. 
They have been widely explored in compact devices such as mobile phones, gloves,buttons
~\cite{brown2006multidimensional, brewster2004tactons, park2020augmenting}, and furniture~\cite{Emgin_2019}.
Beyond simple cues, vibration-based techniques can also render surface textures using high-frequency acoustic signals that modulate friction during user motion~\cite{bau2010teslatouch, winfield2007t}. 
These approaches demonstrate the expressiveness of vibrotactile feedback, particularly on rigid and well-controlled surfaces.
However, scaling vibrotactile feedback to large ambient surfaces typically requires dense arrays of actuators~\cite{Emgin_2019, tsujita2025haptoroom}, which proportionally increases fabrication complexity, and integration overhead. 
These requirements limit the practicality of vibration-based approaches for unobtrusive and ubiquitous integration into everyday objects.

\paragraph{Shape-Changing Haptic Displays}\label{sec:related-2}
Shape-changing displays modulate the surface's distributed geometry, including height and curvature, to generate distinct feedback.
These methods utilize a densely packed array of motor pins~\cite{follmer2013inform, siu2019shapecad} or electrically responsive materials~\cite{coelho2008surflex,besse2017flexible}. 
However, both methods, either a dense actuator array or electrically responsive materials such as shape memory alloys and electroactive polymers, require high power, which makes them hard to scale for embedding in ambient surfaces.

\paragraph{Pressure-Based Tactile Feedback }
Another line of work leverages changes in pressure sensation on the skin to generate distinct tactile feedback.
These works use large air compressors or pumps to inflate the actuation zone in a controlled manner to enable interaction \cite{gunther2020pneumovolley,vazquez20153d}.
To increase feedback diversity, a subcategory of work uses a pneumatic array \cite {shultz2023flat,shen2023fluid} or a phase-changing material \cite{9050916}.
This set of works is highly effective for use in Virtual Reality or gaming interactions.
However, the need for large air compressors or high-power vacuum motors makes these solutions unsuitable for use in ambient environments. 

\paragraph{Kinesthetic Interfaces}
Kinesthetic interfaces provide force-based user feedback through forces, torques, position constraints, and joint or limb impedance \cite{fang2020wireality, achberger2021strive}.
These joints allow for higher degrees of freedom (DoF) \cite{evreinova2014kinesthetic}, so that they can more accurately map task-space interactions to the human body \cite{10.1145/3173574.3174034}, improving transparency and allowing users to perceive and respond to complex interaction forces \cite{10.1145/2984511.2984526, jadhav2017soft}.
As a result, kinesthetic interfaces are particularly effective for tasks requiring precise manipulation and physically grounded interaction, such as virtual object handling and teleoperation \cite{evreinova2014kinesthetic, hannaford1989design}.
However, because these systems require additional actuators to achieve this feedback, the hardware form factor increases \cite{8540015}.
This limitation decreases their usability as an unobtrusive feedback system.
Therefore, we explore methods to reduce the need for additional actuators and hardware.


\subsection{Tangible Origami Interfaces}\label{sec:related-3}
Origami, meaning folding (ori) paper (kami), is an art form that originates in Japan \cite{hatori2011history}.
The structural properties of origami folds have been explored in many fields, such as structural engineering, space exploration, sensor construction, etc. \cite{yue2023review,rohal2024don,muller2024origami}
The Origami-inspired design is widely adopted in HCI.
For example, MiuraKit \cite{cui2023miurakit}, Foldio \cite{olberding2015foldio} and PaperTouch \cite{ye2024papertouch} to enable tangible object interactions.
While works like PneUI \cite{yao2013pneui}, Miurakit \cite{cui2023miurakit} and aerMorph \cite{ou2016aeromorph}
uses the soft shape changing material along with pneumatic pumps to generate programmable structures, which uses airflow to generate desired shape or folding.
However, reliance on pneumatic pumps increases their power requirement and make them less scalable.
The other line of work, such as Printed Paper Actuator\cite{10.1145/3173574.3174143} uses paper for actuation along with the printed circuit.
All these origami-related works are sufficient for application-specific contexts, such as a foldable feedback system \cite{10.1145/3592411}, human tactile peripherals \cite{10.1145/3313831.3376655}, or robotics \cite{10.1145/3706598.3714282, Mintchev2019Dec}.
For \name to adapt an origami folding design into an adaptable actuation mechanism, we address how the folding mechanism enables pressure-based tactile cues across a variety of surfaces.

\section{\name Design}\label{sec:design}
\name adopts a structure-driven approach to generate tactile feedback. 
Instead of relying on actuators placed beneath each interaction point, \name uses a Miura-ori origami-inspired compliant surface structure that distributes actuation across multiple interaction points on the surface. 
The compliant design transforms the lateral actuation into localized out-of-plane deformation, producing normal force for tactile feedback and decoupling feedback generation from actuator placement.

\begin{figure}
    \centering
    \includegraphics[width=0.9\linewidth]{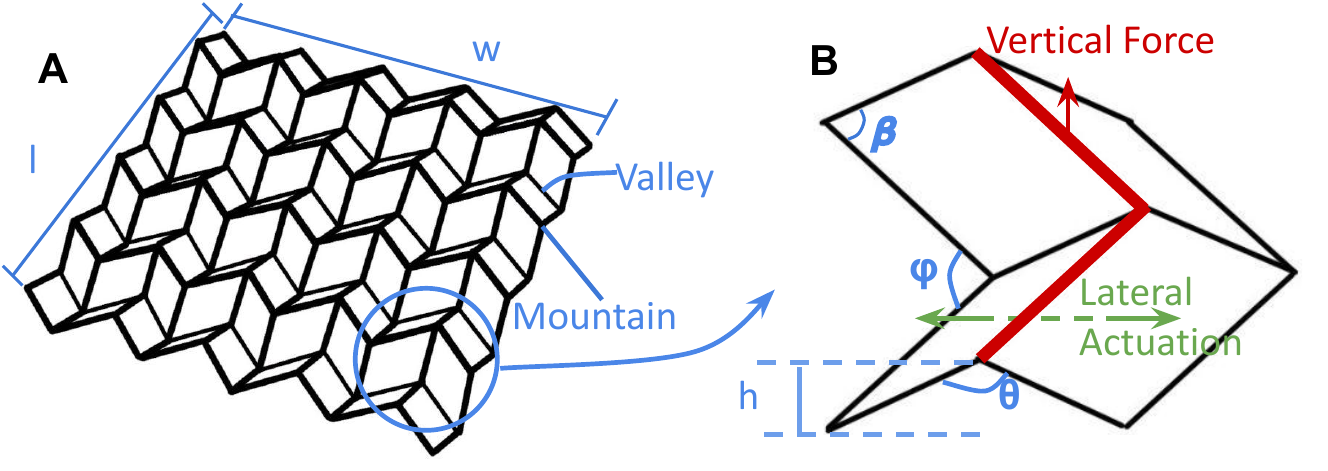}
    \caption{Miura-ori fold (A) and a unit cell (B).}
    \label{fig:miura-ori_fold}
    \vspace{-1ex}
\end{figure}

\begin{figure}
    \centering
    \includegraphics[width=0.9\linewidth]{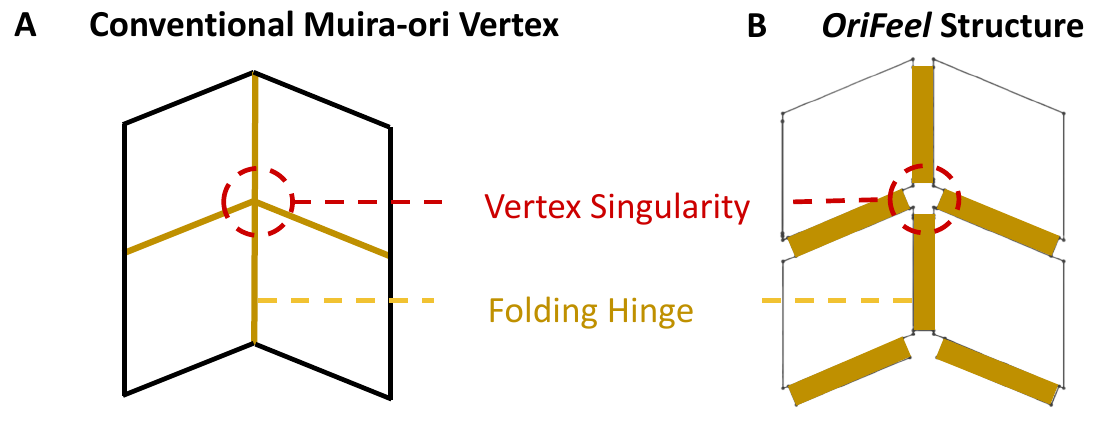}
    \caption{Structural adaptation for Miura-ori origami. (A) The paper origami with yellow folding hinges and a vertex singularity in the red region; (B) the adapted structure for controlled origami folding. }
   \label{fig:structure_drawing}
   \vspace{-3ex}
\end{figure}

\subsection{Miura-ori Structure for Decoupling Actuation from Feedback Generation}
We select the Miura-ori origami pattern \cite{miura2009science} as the basis for our structure-driven design for two reasons: (1) it transforms lateral actuation into normal surface deformation, and (2) its tessellated pattern enables coverage over a large surface and coordinated deformation across multiple actuation points.
Figure \ref{fig:miura-ori_fold}A depicts the tessellated structure of the Miura-ori origami \cite{miura2009science}, while Figure \ref{fig:miura-ori_fold}B shows the unit cell of the foldable pattern.
The structure consists of repeated unit cells that transform lateral actuation into out-of-plane normal displacement, enabling the physical decoupling of actuation from feedback generation.
When lateral actuation (marked by green arrows in Figure \ref{fig:miura-ori_fold}B) is applied, the coupled folding motion of the unit cells produces a vertical force (marked by red arrows in Figure \ref{fig:miura-ori_fold}B), which is perceived as normal-pressure tactile feedback.

\begin{figure*}
    \centering
    \begin{minipage}{0.3\textwidth}
        \centering
            \includegraphics[width=\linewidth]{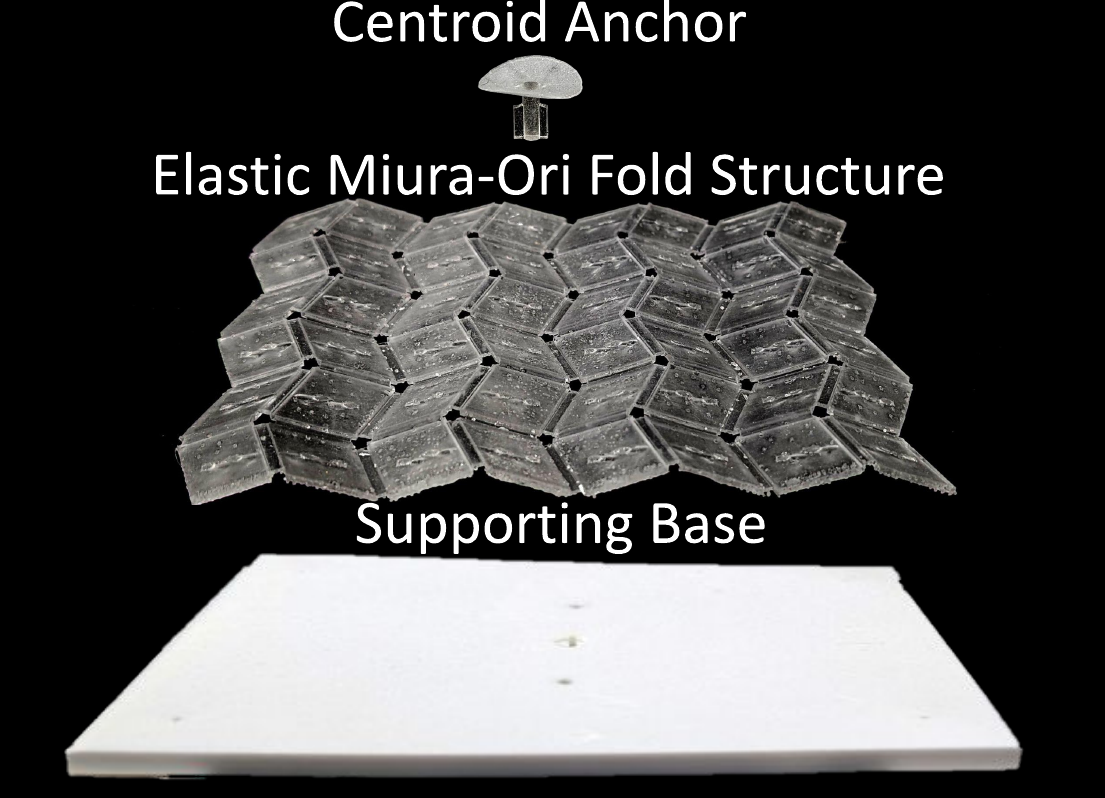}
            \caption{\name prototype, including key components: centroid anchor, folding structure, base plate for routing embedded cables to actuators.}
    \label{fig:layers}
    \end{minipage}
     \hfill
    \begin{minipage}{0.27\textwidth}
        \centering
            \includegraphics[width=\linewidth]{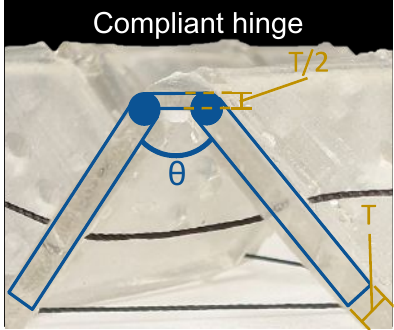}
            \caption{Compliant design, two facets of thickness $T$ connected by thin hinge of thickness $\frac{T}{2}$ with embedded cable for actuation.} 
        \label{fig:compliant}
    \end{minipage}
    \hfill
    \begin{minipage}{0.3\textwidth}
        \centering
                        \includegraphics[width=\linewidth]{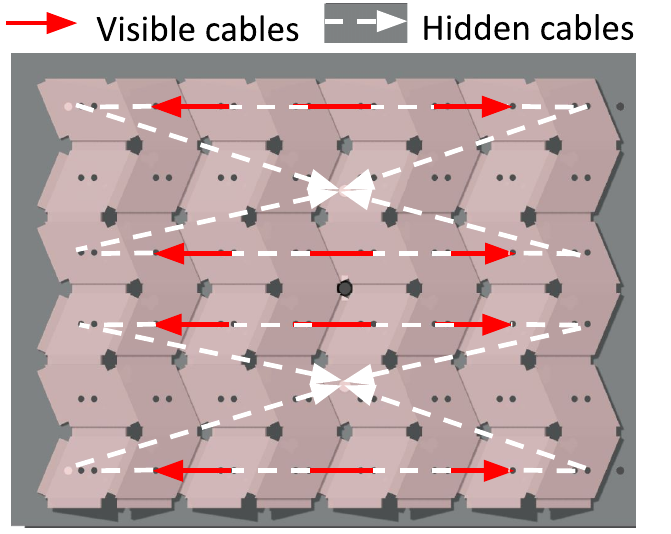}
            \caption{Cable-based actuation design. Arrows show the pull force along the cable when actuated, red lines indicating visible lines from top view.}
    \label{fig:structure_control}
    \end{minipage}
\end{figure*}
\subsection{Compliant Structure Adaptation for Ubiquitous Surfaces}\label{sec:compliant}
While the Miura-ori mechanism provides desired folding kinematic properties, integrating it into ambient surfaces still requires controlled, stable folding in non-ideal materials.
However, most materials either resist bending~\cite{morris2017strength} or deform elastically without retaining a folded state~\cite{bensonbending}, making controlled folding difficult.
Figure~\ref{fig:structure_drawing}A shows the top view of a conventional Miura-ori fold vertex, where multiple creases meet at a single point, resulting in tightly coupled deformation that limits the degree of controllable folding in non-ideal materials.

To address these challenges, we adapt the structural design in two ways.
First, \name adopts a compliant hinge design that introduces localized reduced thickness, creating regions that undergo controlled folding under actuation—effectively mimicking paper-like creases in non-paper materials~\cite{wagner2020hinges, meloni2021engineering}.
This design localizes deformation by concentrating bending in the thin hinge regions while the thick facet remains comparatively rigid
\cite{howell2013introduction, lobontiu2002compliant}.
As illustrated in Figure~\ref{fig:structure_drawing}B, we implement the folding lines as thin flexure hinges with reduced thickness relative to adjacent facets, enabling controlled and repeatable folding.
To alleviate the over-constrained deformation at the vertex, we introduce a small gap at the intersection, breaking the direct coupling between adjacent facets. This relaxes the geometric constraint imposed by the shared vertex, allowing the fold lines to deform more independently, and enabling stable, controllable folding in non-ideal materials.

\subsection{Prototype Implementation}
To demonstrate the effectiveness of structure-driven actuation design, we implemented \name with both rigid and soft materials.
The Miura-ori-inspired folding structure is divided into two parts: (1) the parallelogram facet, (2) the folding hinge.
Four facets are connected to create a single folding structural unit.
Two neighboring facets are connected via a folding-compliant hinge; to concentrate the actuation at the hinge, both the facet thickness and the hinge thickness are $2$ mm and $1$ mm, respectively.
The thickness difference concentrates actuation on the thin folding hinge, generating the folding effect.
Each side of the facet is kept at $22$ mm, and the folding hinge is $18$ mm long and $3$ mm wide.
This difference in length creates the empty-vertex singularity region, as shown in Figure \ref{fig:structure_drawing}B, to accommodate smooth folding.

The rigid structure is fabricated using Polylactic acid (PLA)  plastic filament on a Fused Deposition Modeling (FDM) printer, while the elastic structure is fabricated using Formlabs 50A resin \cite{Formlabs_Elastic50A_Resin} on the Formlabs 4B printer \cite{Formlabs_Form4B_2024}.
\name embeds cables across the folding units to distribute actuation from a remote actuator that can be placed flexibly.
To accommodate string embedding, each facet is printed with two 2 mm cable-embedding holes at its center.
The printed foldable module consists of a $3 \times 4$ unit pattern. 
We utilize origami's limited degree of freedom to distribute the actuation force across multiple folding units. 
A total of four cables are embedded in the structure through the center holes, with one cable in the first row, two cables in the second row, and one cable in the third row, as shown in Figure \ref{fig:layers}.
This configuration of embedded cables allows \name to generate distributed tactile forces across the units.



Figure \ref{fig:layers} shows different components of the \name prototype.
The centroid keeps the folding structure stable around a center point to prevent drifting during the actuation process.
The centroid is followed by an elastic Miura-ori fold structure with compliant hinges, shown separately in Figure \ref{fig:compliant}.
The centroid and the elastic fold structure sit on top of the supporting base plate.
The centroid anchor connects the folding structure to the supporting base.
The base also provides a path to route the cables connected to the actuator.
Figure \ref{fig:structure_control} shows how cables are embedded in the folding structure to control multiple folding units.
Although the paper origami has a single degree of freedom, the structure behaves differently.
Therefore, the \name prototype embeds multiple cables to further distribute the actuation across the whole structure.

\begin{figure}
    \centering
    \includegraphics[width=\linewidth]{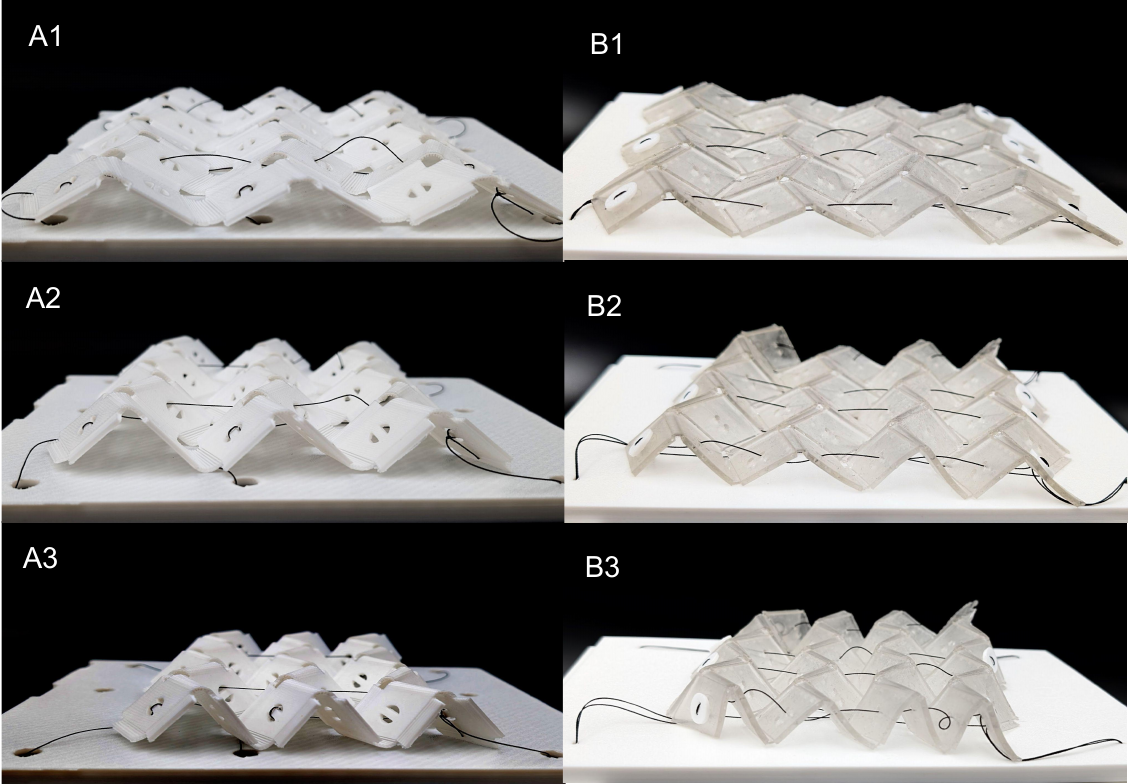}
    \caption{Different actuation configurations of \name structures with (A) rigid material and (B) elastic material}
    \label{fig:folding_results}
    \vspace{-2ex}
\end{figure}


\section{Mechanical Characterization}\label{sec:structure-test}
We evaluated how \name distributes actuation from a single actuator across multiple units to achieve a uniform folding effect in both rigid and elastic implementations.
We actuate the folding prototype at three different actuation levels (level 1, level 2, and level 3) based on the cable pull.
For each input actuation, the folding structure is allowed to reach mechanical equilibrium without external loads.
We capture high-resolution lateral images of the folding structure and perform an optical calculation of the folding angle $\theta$.
Figure \ref{fig:folding_results} shows two hardware configurations: A) a rigid prototype, B) an elastic prototype, and three different actuation levels.
Table \ref{tab:folding_error} summarizes the mean fold angle $\theta$ and standard deviation measured across the actuated units.
We observe that across all three actuation levels, both models exhibit similar fold angles.
The rigid model A shows more consistent folding than the elastic model B across three actuation levels.
This is because, in the rigid model A, the facets remain comparatively rigid and concentrate deformation in the compliant hinge.
For the elastic model B, a portion of the actuation deforms the facet itself, resulting in more inconsistent folding with a higher standard deviation compared to the rigid model A.
The overall standard deviation is low, which indicates that embedded cables produce coordinated deformation across all units rather than localized folding.

\begin{table}[!t]
    \centering
    \caption{Average fold angles in degrees for different actuation configurations for rigid structure A and soft structure B. }
    \label{tab:folding_error}
    \begin{tabular}{|c|c|c|c|c|}
    \hline
      \textbf{Config}   & \textbf{Mean A } & \textbf{STD A}& \textbf{Mean B}& \textbf{STD B} \\
          \hline    \hline
       \textbf{$1$}  & 132.77 &2.71 & 118.97&13.20\\
       \textbf{$2$}  & 89.66 & 1.52& 95.32&12.39\\
       \textbf{$3$}  & 67.72 & 1.56& 55.92&19.10\\
           \hline
    \end{tabular}
    \vspace{-4ex}
\end{table}

\section{Discussion and Future Direction}\label{sec:discussion}
Our results show that \name's structure-driven approach enables a compact actuator footprint with one-to-many actuation.
We envision several future directions to be explored to realize the full potential of this folding-based surface interface.

\paragraph{Expressive Haptic Feedback Control}
This work demonstrates the feasibility of one-to-many actuation with uniformly distributed tactile output.
However, the true potential of this design lies in its interconnected structure, which enables the generation of complex spatial tactile patterns with fewer actuators.
In future work, we plan to explore the rich control space afforded by the interconnected origami structure by systematically characterizing the trade-offs between haptic expressiveness and actuator configuration, including actuator count and placement.

\paragraph{Large Scale Integration}
This work demonstrates the feasibility of the compact actuator design for surface folding. 
Although it is an important step towards ubiquitous haptic interfaces, there are still challenges in integrating the design into large surfaces on everyday objects.
Actuating larger interconnected structures may introduce higher friction, increased force attenuation, and reduced control precision across the surface.
Therefore, it is essential to characterize the achievable control space and identify the set of perceptually distinguishable haptic interactions under different actuator configurations and structural designs. 
Such studies of the trade-offs between control precision and surface scale will establish design guidelines for balancing haptic expressiveness, actuator density, and scalability in large-area ambient haptic surfaces.

\paragraph{Perceptual Evaluation with Users}
The evaluation in this work focuses on the mechanical characterization of the folding structure and its objective capability to generate pressure-based tactile feedback. 
However, the practical utility of such interfaces ultimately depends on how users perceive this feedback across diverse populations and usage conditions. 
Our future work includes investigating perceptual thresholds, discrimination accuracy, and subjective experience across diverse user populations—including older adults and individuals with varying levels of tactile sensitivity—as well as across different body locations that commonly interact with ambient surfaces.
Such studies will establish the perceptual limits and configuration guidelines of the interactive folding surface design, identify user-specific design considerations, and inform the development of accessible haptic interfaces for everyday ambient surfaces.

\section{Conclusion}\label{sec:conclusion}
This paper presented \name, a structure-driven tactile interface that rethinks haptic feedback generation by transforming the surface itself into an active medium for tactile interaction via folding. 
By combining compliant Miura-Ori-inspired folding with decoupled actuation, \name enables compact actuator integration, controlled surface deformation, and one-to-many tactile actuation. 
Through prototype implementation and mechanical characterization, we demonstrated the feasibility of distributed normal-pressure feedback generated through structural interconnectivity. 
We envision this work as a first step toward scalable, unobtrusive haptic interfaces that seamlessly integrate into the everyday ambient surfaces surrounding us.

\bibliographystyle{ACM-Reference-Format}
\bibliography{ref}

@article{meloni2021engineering,
  title={Engineering origami: A comprehensive review of recent applications, design methods, and tools},
  author={Meloni, Marco and Cai, Jianguo and Zhang, Qian and Sang-Hoon Lee, Daniel and Li, Meng and Ma, Ruijun and Parashkevov, Teo Emilov and Feng, Jian},
  journal={Advanced Science},
  volume={8},
  number={13},
  pages={2000636},
  year={2021},
  publisher={Wiley Online Library}
}

@article{wagner2020hinges,
  title={Hinges for origami-inspired structures by multimaterial additive manufacturing},
  author={Wagner, Marius A and Huang, Jian-Lin and Okle, Philipp and Paik, Jamie and Spolenak, Ralph},
  journal={Materials \& Design},
  volume={191},
  pages={108643},
  year={2020},
  publisher={Elsevier}
}

@misc{bensonbending,
  title={Bending Basics: The hows and whys of springback and springforward, 2014},
  author={Benson, S}
}

@article{morris2017strength,
  title={Strength, Stiffness, and Abuse Resistance},
  author={Morris, BA},
  journal={The Science and Technology of Flexible Packaging; Elsevier: Oxford, UK},
  pages={309--350},
  year={2017}
}

@inproceedings{shultz2023flat,
  title={Flat panel haptics: Embedded electroosmotic pumps for scalable shape displays},
  author={Shultz, Craig and Harrison, Chris},
  booktitle={Proceedings of the 2023 CHI conference on human factors in computing systems},
  pages={1--16},
  year={2023}
}

@inproceedings{yao2013pneui,
  title={PneUI: pneumatically actuated soft composite materials for shape changing interfaces},
  author={Yao, Lining and Niiyama, Ryuma and Ou, Jifei and Follmer, Sean and Della Silva, Clark and Ishii, Hiroshi},
  booktitle={Proceedings of the 26th annual ACM symposium on User interface software and Technology},
  pages={13--22},
  year={2013}
}

@inproceedings{ou2016aeromorph,
  title={aeroMorph-heat-sealing inflatable shape-change materials for interaction design},
  author={Ou, Jifei and Skouras, M{\'e}lina and Vlavianos, Nikolaos and Heibeck, Felix and Cheng, Chin-Yi and Peters, Jannik and Ishii, Hiroshi},
  booktitle={Proceedings of the 29th Annual Symposium on User Interface Software and Technology},
  pages={121--132},
  year={2016}
}

@inproceedings{olberding2015foldio,
  title={Foldio: Digital fabrication of interactive and shape-changing objects with foldable printed electronics},
  author={Olberding, Simon and Soto Ortega, Sergio and Hildebrandt, Klaus and Steimle, J{\"u}rgen},
  booktitle={Proceedings of the 28th Annual ACM Symposium on User Interface Software \& Technology},
  pages={223--232},
  year={2015}
}

@inproceedings{cui2023miurakit,
  title={MiuraKit: A Modular Hands-On Construction Kit For Pneumatic Shape-Changing And Robotic Interfaces},
  author={Cui, Zhitong and Wang, Shuhong and Li, Junxian and Luo, Shijian and Ion, Alexandra},
  booktitle={Proceedings of the 2023 ACM Designing Interactive Systems Conference},
  pages={2066--2078},
  year={2023}
}

@inproceedings{ye2024papertouch,
  title={PaperTouch: Tangible Interfaces through Paper Craft and Touchscreen Devices},
  author={Ye, Qian and Yong, Zhen Zhou and Han, Bo and Yen, Ching Chiuan and Zheng, Clement},
  booktitle={Proceedings of the 2024 CHI Conference on Human Factors in Computing Systems},
  year={2024}
}

@incollection{coelho2008surflex,
  title={Surflex: a programmable surface for the design of tangible interfaces},
  author={Coelho, Marcelo and Ishii, Hiroshi and Maes, Pattie},
  booktitle={CHI'08 extended abstracts on Human factors in computing systems},
  pages={3429--3434},
  year={2008}
}

@article{besse2017flexible,
  title={Flexible active skin: large reconfigurable arrays of individually addressed shape memory polymer actuators},
  author={Besse, Nadine and Rosset, Samuel and Zarate, Juan Jose and Shea, Herbert},
  journal={Advanced Materials Technologies},
  volume={2},
  number={10},
  pages={1700102},
  year={2017},
  publisher={Wiley Online Library}
}

@inproceedings{bau2010teslatouch,
  title={TeslaTouch: electrovibration for touch surfaces},
  author={Bau, Olivier and Poupyrev, Ivan and Israr, Ali and Harrison, Chris},
  booktitle={Proceedings of the 23nd annual ACM symposium on User interface software and technology},
  pages={283--292},
  year={2010}
}

@inproceedings{winfield2007t,
  title={T-pad: Tactile pattern display through variable friction reduction},
  author={Winfield, Laura and Glassmire, John and Colgate, J Edward and Peshkin, Michael},
  booktitle={Second Joint EuroHaptics Conference and Symposium on Haptic Interfaces for Virtual Environment and Teleoperator Systems (WHC'07)},
  pages={421--426},
  year={2007},
  organization={IEEE}
}

@inproceedings{follmer2013inform,
  title={inFORM: dynamic physical affordances and constraints through shape and object actuation.},
  author={Follmer, Sean and Leithinger, Daniel and Olwal, Alex and Hogge, Akimitsu and Ishii, Hiroshi},
  booktitle={Uist},
  volume={13},
  number={10},
  pages={2501--988},
  year={2013}
}

@book{lobontiu2002compliant,
  title={Compliant mechanisms: design of flexure hinges},
  author={Lobontiu, Nicolae},
  year={2002},
  publisher={ }
}

@misc{Formlabs_Form4B_2024,
  title        = {Formlabs Form 4B 3D Printer},
  author       = {{Formlabs}},
  year         = {2024},
  howpublished = {\url{https://formlabs.com/3d-printers/form-4b/}}
}

@misc{Formlabs_Elastic50A_Resin,
  title        = {Elastic 50A Resin},
  author       = {{Formlabs}},
  year         = {2024},
  howpublished = {\url{https://formlabs.com/store/materials/elastic-50a-resin-v2/}}
}

@inproceedings{siu2019shapecad,
  title={shapeCAD: An accessible 3D modelling workflow for the blind and visually-impaired via 2.5 D shape displays},
  author={Siu, Alexa F and Kim, Son and Miele, Joshua A and Follmer, Sean},
  booktitle={Proceedings of the 21st International ACM SIGACCESS Conference on Computers and Accessibility},
  pages={342--354},
  year={2019}
}

@article{howell2013introduction,
  title={Introduction to compliant mechanisms},
  author={Howell, Larry L},
  journal={Handbook of Compliant Mechanisms},
  pages={1--13},
  year={2013},
  publisher={Wiley Online Library}
}

@inproceedings{vazquez20153d,
  title={3d printing pneumatic device controls with variable activation force capabilities},
  author={V{\'a}zquez, Marynel and Brockmeyer, Eric and Desai, Ruta and Harrison, Chris and Hudson, Scott E},
  booktitle={Proceedings of the 33rd Annual ACM Conference on Human Factors in Computing Systems},
  pages={1295--1304},
  year={2015}
}

@inproceedings{10.1145/3173574.3174143,
author = {Wang, Guanyun and Cheng, Tingyu and Do, Youngwook and Yang, Humphrey and Tao, Ye and Gu, Jianzhe and An, Byoungkwon and Yao, Lining},
title = {Printed Paper Actuator: A Low-cost Reversible Actuation and Sensing Method for Shape Changing Interfaces},
year = {2018},
isbn = {9781450356206},
publisher = {Association for Computing Machinery},
address = {New York, NY, USA},
url = {https://doi.org/10.1145/3173574.3174143},
doi = {10.1145/3173574.3174143},
booktitle = {Proceedings of the 2018 CHI Conference on Human Factors in Computing Systems},
pages = {1–12},
numpages = {12},
location = {Montreal QC, Canada},
series = {CHI '18}
}

@article{hatori2011history,
  title={History of Origami in the East and the West before Interfusion},
  author={Hatori, Koshiro},
  journal={Origami},
  volume={5},
  pages={3--11},
  year={2011},
  publisher={CRC Press}
}

@inproceedings{yue2023review,
  title={A review of origami-based deployable structures in aerospace engineering},
  author={Yue, Songlin},
  booktitle={Journal of Physics: Conference Series},
  volume={2459},
  number={1},
  pages={012137},
  year={2023}
}

@article{muller2024origami,
  title={origami housing: An innovative and resilient postdisaster temporary emergency housing solution},
  author={M{\"u}ller, Claudia Calle and ElZomor, Mohamed},
  journal={Journal of Architectural Engineering},
  volume={30},
  number={3},
  pages={04024025},
  year={2024},
  publisher={American Society of Civil Engineers}
}

@inproceedings{rohal2024don,
  title={Don’t Crosstalk to Me: Origami Structure-Augmented Sensing for Scalable Surface Pressure Monitoring},
  author={Rohal, Shubham and Lee, Dong Yoon and Ruiz, Carlos and Zhang, Joshua and Fagert, Jonathon and Han, Jun and Pan, Shijia},
  booktitle={Proceedings of the 22nd ACM Conference on Embedded Networked Sensor Systems},
  pages={493--506},
  year={2024}
}

@article{miura2009science,
  title={The science of Miura-ori: A review},
  author={Miura, Koryo and Lang, Robert J},
  journal={Origami},
  volume={4},
  number={3653},
  pages={87},
  year={2009},
  publisher={Kerswell Books}
}

@inproceedings{fang2020wireality,
  title={Wireality: Enabling complex tangible geometries in virtual reality with worn multi-string haptics},
  author={Fang, Cathy and Zhang, Yang and Dworman, Matthew and Harrison, Chris},
  booktitle={Proceedings of the 2020 CHI Conference on Human Factors in Computing Systems},
  pages={1--10},
  year={2020}
}

@inproceedings{achberger2021strive,
  title={Strive: String-based force feedback for automotive engineering},
  author={Achberger, Alexander and Aust, Fabian and Pohlandt, Daniel and Vidackovic, Kresimir and Sedlmair, Michael},
  booktitle={The 34th Annual ACM Symposium on User Interface Software and Technology},
  pages={841--853},
  year={2021}
}

@inproceedings{park2020augmenting,
  title={Augmenting physical buttons with vibrotactile feedback for programmable feels},
  author={Park, Chaeyong and Yoon, Jinhyuk and Oh, Seungjae and Choi, Seungmoon},
  booktitle={Proceedings of the 33rd Annual ACM Symposium on User Interface Software and Technology},
  pages={924--937},
  year={2020}
}

@article{Emgin_2019,
   title={HapTable: An Interactive Tabletop Providing Online Haptic Feedback for Touch Gestures},
   volume={25},
   ISSN={2160-9306},
   url={http://dx.doi.org/10.1109/TVCG.2018.2855154},
   DOI={10.1109/tvcg.2018.2855154},
   number={9},
   journal={IEEE Transactions on Visualization and Computer Graphics},
   publisher={Institute of Electrical and Electronics Engineers (IEEE)},
   author={Emgin, Senem Ezgi and Aghakhani, Amirreza and Sezgin, T. Metin and Basdogan, Cagatay},
   year={2019},
   month=Sept, pages={2749–2762} }

@inproceedings{tsujita2025haptoroom,
  title={Haptoroom: Designing spatial and reconfigurable haptic experience using vibrotactile floor interface},
  author={Tsujita, Kiryu and Yoshida, Takatoshi and Kobayashi, Kohei and Hanamitsu, Nobuhisa and Zhong, Kanghong and Horie, Arata and Minamizawa, Kouta},
  booktitle={Proceedings of the Nineteenth International Conference on Tangible, Embedded, and Embodied Interaction},
  pages={1--11},
  year={2025}
}

@inproceedings{10.1145/2984511.2984526,
author = {Benko, Hrvoje and Holz, Christian and Sinclair, Mike and Ofek, Eyal},
title = {NormalTouch and TextureTouch: High-fidelity 3D Haptic Shape Rendering on Handheld Virtual Reality Controllers},
year = {2016},
isbn = {9781450341899},
publisher = {Association for Computing Machinery},
address = {New York, NY, USA},
url = {https://doi.org/10.1145/2984511.2984526},
doi = {10.1145/2984511.2984526},
booktitle = {Proceedings of the 29th Annual Symposium on User Interface Software and Technology},
pages = {717–728},
numpages = {12},
location = {Tokyo, Japan},
series = {UIST '16}
}

@article{hannaford1989design,
  title={A design framework for teleoperators with kinesthetic feedback},
  author={Hannaford, Blake},
  journal={IEEE transactions on Robotics and Automation},
  volume={5},
  number={4},
  pages={426--434},
  year={1989},
  publisher={IEEE}
}

@article{jadhav2017soft,
  title={Soft robotic glove for kinesthetic haptic feedback in virtual reality environments},
  author={Jadhav, Saurabh and Kannanda, Vikas and Kang, Bocheng and Tolley, Michael T and Schulze, Jurgen P},
  journal={electronic imaging},
  volume={29},
  pages={19--24},
  year={2017},
  publisher={Society for Imaging Science and Technology}
}

@article{evreinova2014kinesthetic,
  title={From kinesthetic sense to new interaction concepts: Feasibility and constraints},
  author={Evreinova, Tatiana V and Evreinov, Grigori and Raisamo, Roope},
  journal={International Journal of Advanced Computer Technology},
  volume={3},
  number={4},
  pages={1--33},
  year={2014}
}

@inproceedings{10.1145/3173574.3174034,
author = {Rietzler, Michael and Geiselhart, Florian and Frommel, Julian and Rukzio, Enrico},
title = {Conveying the Perception of Kinesthetic Feedback in Virtual Reality using State-of-the-Art Hardware},
year = {2018},
isbn = {9781450356206},
publisher = {Association for Computing Machinery},
address = {New York, NY, USA},
url = {https://doi.org/10.1145/3173574.3174034},
doi = {10.1145/3173574.3174034},
booktitle = {Proceedings of the 2018 CHI Conference on Human Factors in Computing Systems},
pages = {1–13},
numpages = {13},
location = {Montreal QC, Canada},
series = {CHI '18}
}

@inproceedings{gunther2020pneumovolley,
  title={PneumoVolley: Pressure-based haptic feedback on the head through pneumatic actuation},
  author={G{\"u}nther, Sebastian and Sch{\"o}n, Dominik and M{\"u}ller, Florian and M{\"u}hlh{\"a}user, Max and Schmitz, Martin},
  booktitle={Extended abstracts of the 2020 CHI conference on human factors in computing systems},
  pages={1--10},
  year={2020}
}

@ARTICLE{8540015,
  author={Wang, Dangxiao and Song, Meng and Naqash, Afzal and Zheng, Yukai and Xu, Weiliang and Zhang, Yuru},
  journal={IEEE Transactions on Haptics}, 
  title={Toward Whole-Hand Kinesthetic Feedback: A Survey of Force Feedback Gloves}, 
  year={2019},
  volume={12},
  number={2},
  pages={189-204},
  doi={10.1109/TOH.2018.2879812}}

@inproceedings{shen2023fluid,
  title={Fluid reality: High-resolution, untethered haptic gloves using electroosmotic pump arrays},
  author={Shen, Vivian and Rae-Grant, Tucker and Mullenbach, Joe and Harrison, Chris and Shultz, Craig},
  booktitle={Proceedings of the 36th Annual ACM Symposium on User Interface Software and Technology},
  pages={1--20},
  year={2023}
}

@inproceedings{brown2006multidimensional,
  title={Multidimensional tactons for non-visual information presentation in mobile devices},
  author={Brown, Lorna M and Brewster, Stephen A and Purchase, Helen C},
  booktitle={Proceedings of the 8th conference on Human-computer interaction with mobile devices and services},
  pages={231--238},
  year={2006}
}

@article{brewster2004tactons,
  title={Tactons: structured tactile messages for non-visual information display},
  author={Brewster, Stephen A and Brown, Lorna M},
  year={2004},
  publisher={Australian Computer Society}
}

@inproceedings{10.1145/3706598.3714282,
author = {Li, Jiaji and Feng, Shuyue and Perroni-Scharf, Maxine and Liu, Yujia and Guan, Emily and Wang, Guanyun and Mueller, Stefanie},
title = {Xstrings: 3D Printing Cable-Driven Mechanism for Actuation, Deformation, and Manipulation},
year = {2025},
isbn = {9798400713941},
publisher = {Association for Computing Machinery},
address = {New York, NY, USA},
url = {https://doi.org/10.1145/3706598.3714282},
doi = {10.1145/3706598.3714282},
booktitle = {Proceedings of the 2025 CHI Conference on Human Factors in Computing Systems},
articleno = {6},
numpages = {17},
location = {
},
series = {CHI '25}
}

@article{Mintchev2019Dec,
	author = {Mintchev, Stefano and Salerno, Marco and Cherpillod, Alexandre and Scaduto, Simone and Paik, Jamie},
	title = {{A portable three-degrees-of-freedom force feedback origami robot for human{\textendash}robot interactions}},
	journal = {Nat. Mach. Intell.},
	volume = {1},
	pages = {584--593},
	year = {2019},
	month = dec,
	issn = {2522-5839},
	publisher = {Nature Publishing Group},
	doi = {10.1038/s42256-019-0125-1}
}

@inproceedings{10.1145/3313831.3376655,
author = {Chang, Zekun and Ta, Tung D. and Narumi, Koya and Kim, Heeju and Okuya, Fuminori and Li, Dongchi and Kato, Kunihiro and Qi, Jie and Miyamoto, Yoshinobu and Saito, Kazuya and Kawahara, Yoshihiro},
title = {Kirigami Haptic Swatches: Design Methods for Cut-and-Fold Haptic Feedback Mechanisms},
year = {2020},
isbn = {9781450367080},
publisher = {Association for Computing Machinery},
address = {New York, NY, USA},
url = {https://doi.org/10.1145/3313831.3376655},
doi = {10.1145/3313831.3376655},
booktitle = {Proceedings of the 2020 CHI Conference on Human Factors in Computing Systems},
pages = {1–12},
numpages = {12},
location = {Honolulu, HI, USA},
series = {CHI '20}
}

@article{10.1145/3592411,
author = {Freire, Marco and Bhargava, Manas and Schreck, Camille and Hugron, Pierre-Alexandre and Bickel, Bernd and Lefebvre, Sylvain},
title = {PCBend: Light Up Your 3D Shapes With Foldable Circuit Boards},
year = {2023},
issue_date = {August 2023},
publisher = {Association for Computing Machinery},
address = {New York, NY, USA},
volume = {42},
number = {4},
issn = {0730-0301},
url = {https://doi.org/10.1145/3592411},
doi = {10.1145/3592411},
journal = {ACM Trans. Graph.},
month = jul,
articleno = {142},
numpages = {16}
}

@ARTICLE{9050916,
  author={Narumi, Koya and Sato, Hiroki and Nakahara, Kenichi and Seong, Young ah and Morinaga, Kunihiko and Kakehi, Yasuaki and Niiyama, Ryuma and Kawahara, Yoshihiro},
  journal={IEEE Robotics and Automation Letters}, 
  title={Liquid Pouch Motors: Printable Planar Actuators Driven by Liquid-to-Gas Phase Change for Shape-Changing Interfaces}, 
  year={2020},
  volume={5},
  number={3},
  pages={3915-3922},
  doi={10.1109/LRA.2020.2983681}}

\end{document}